\documentclass[twocolumn]{aastex63}
\usepackage{graphicx}
\usepackage{subfigure}
\usepackage{color, epsfig}
\usepackage{apjfonts, natbib}
\usepackage{appendix}
\usepackage{float}
\usepackage{bm}
\usepackage{hyperref}
\usepackage{lineno}

\newcommand{\hb}{H\ensuremath{\beta}}
\newcommand{\lum}{erg\,s\ensuremath{^{-1}}}

\newcommand{\mgii}{Mg\,{\footnotesize II}}

\newcommand{\msun}{\ensuremath{M_{\odot}}}

\newcommand{\mbh}{\ensuremath{M_\mathrm{BH}}}

\shorttitle{AT~2025wet}
\shortauthors{ZHU et al.}

\begin{document}

\title{\Large Discovery of a Featureless Tidal Disruption Event at $z\sim1$ with the Wide Field Survey Telescope}

\author[0000-0003-3824-9496]{Jiazheng Zhu}
\affiliation{Department of Astronomy, University of Science and Technology of China, Hefei, 230026, China; jiazheng@mail.ustc.edu.cn, jnac@ustc.edu.cn}
\affiliation{School of Astronomy and Space Sciences,
University of Science and Technology of China, Hefei, 230026, China}

\author{Zelin Xu}
\affiliation{Department of Astronomy, University of Science and Technology of China, Hefei, 230026, China; jiazheng@mail.ustc.edu.cn, jnac@ustc.edu.cn}
\affiliation{School of Astronomy and Space Sciences,
University of Science and Technology of China, Hefei, 230026, China}

\author[0000-0002-7152-3621]{Ning Jiang}
\affiliation{Department of Astronomy, University of Science and Technology of China, Hefei, 230026, China; jiazheng@mail.ustc.edu.cn, jnac@ustc.edu.cn}
\affiliation{School of Astronomy and Space Sciences,
University of Science and Technology of China, Hefei, 230026, China}

\author[0000-0002-9092-0593]{Ji-an Jiang}
\affiliation{Department of Astronomy, University of Science and Technology of China, Hefei, 230026, China; jiazheng@mail.ustc.edu.cn, jnac@ustc.edu.cn}
\affiliation{School of Astronomy and Space Sciences,
University of Science and Technology of China, Hefei, 230026, China}

\author[0000-0002-1517-6792]{Tinggui Wang}
\affiliation{Department of Astronomy, University of Science and Technology of China, Hefei, 230026, China; jiazheng@mail.ustc.edu.cn, jnac@ustc.edu.cn}
\affiliation{School of Astronomy and Space Sciences, University of Science and Technology of China, Hefei, 230026, China}
\affiliation{Department of Physics and Astronomy, College of Physics, Guizhou University, Guiyang 550025, People's Republic of China}

\author[0000-0001-6747-8509]{Yuhan Yao}
\affiliation{Department of Astronomy, University of California, Berkeley, CA 94720-3411, USA}
\affiliation{Miller Institute for Basic Research in Science, 206B Stanley Hall, Berkeley, CA 94720, USA
}
\affiliation{Berkeley Center for Multi-messenger Research on Astrophysical Transients and Outreach (Multi-RAPTOR), University of California, Berkeley, CA 94720-3411, USA}

\author[0000-0002-7706-5668]{Ryan Chornock}
\affiliation{Department of Astronomy, University of California, Berkeley, CA 94720-3411, USA}
\affiliation{Berkeley Center for Multi-messenger Research on Astrophysical Transients and Outreach (Multi-RAPTOR), University of California, Berkeley, CA 94720-3411, USA}

\author[0000-0002-5698-8703]{Erica Hammerstein}
\affiliation{Department of Astronomy, University of California, Berkeley, CA 94720-3411, USA}
\affiliation{Berkeley Center for Multi-messenger Research on Astrophysical Transients and Outreach (Multi-RAPTOR), University of California, Berkeley, CA 94720-3411, USA}

\author[0000-0003-4225-5442]{Yibo~Wang}
\affiliation{Department of Astronomy, University of Science and Technology of China, Hefei, 230026, China; jiazheng@mail.ustc.edu.cn, jnac@ustc.edu.cn}
\affiliation{School of Astronomy and Space Sciences, University of Science and Technology of China, Hefei, 230026, China}

\author[0000-0003-4721-6477]{Min-Xuan~Cai}
\affiliation{Department of Astronomy, University of Science and Technology of China, Hefei, 230026, China; jiazheng@mail.ustc.edu.cn, jnac@ustc.edu.cn}
\affiliation{School of Astronomy and Space Sciences,
University of Science and Technology of China, Hefei, 230026, China}

\author[0000-0001-7689-6382]{Shifeng Huang}
\affiliation{Department of Astronomy, University of Science and Technology of China, Hefei, 230026, China; jiazheng@mail.ustc.edu.cn, jnac@ustc.edu.cn}
\affiliation{School of Astronomy and Space Sciences, University of Science and Technology of China, Hefei, 230026, China}

\author[0009-0007-3464-417X]{Wenkai Li}
\affiliation{Department of Astronomy, University of Science and Technology of China, Hefei, 230026, China; jiazheng@mail.ustc.edu.cn, jnac@ustc.edu.cn}
\affiliation{School of Astronomy and Space Sciences, University of Science and Technology of China, Hefei, 230026, China}

\author[0009-0006-8112-0187]{Mingxin Wu}
\affiliation{Department of Astronomy, University of Science and Technology of China, Hefei, 230026, China; jiazheng@mail.ustc.edu.cn, jnac@ustc.edu.cn}
\affiliation{School of Astronomy and Space Sciences,
University of Science and Technology of China, Hefei, 230026, China}

\author[0000-0002-2006-1615]{Chichuan Jin}
\affiliation{National Astronomical Observatories, Chinese Academy of Sciences, Beijing, 100101, China}

\author[0000-0003-3965-6931]{Jie Lin}
\affiliation{Department of Astronomy, University of Science and Technology of China, Hefei, 230026, China; jiazheng@mail.ustc.edu.cn, jnac@ustc.edu.cn}
\affiliation{School of Astronomy and Space Sciences,
University of Science and Technology of China, Hefei, 230026, China}

\author[0000-0002-6221-1829]{Jianwei Lyu}
\affiliation{Steward Observatory, University of Arizona, 933 North Cherry Avenue, Tucson, AZ 85721, USA}

\author{Dezheng Meng}
\affiliation{Department of Astronomy, University of Science and Technology of China, Hefei, 230026, China; jiazheng@mail.ustc.edu.cn, jnac@ustc.edu.cn}
\affiliation{School of Astronomy and Space Sciences,
University of Science and Technology of China, Hefei, 230026, China}

\author[0009-0003-9474-8457]{Weiyu Wu}
\affiliation{Department of Astronomy, University of Science and Technology of China, Hefei, 230026, China; jiazheng@mail.ustc.edu.cn, jnac@ustc.edu.cn}
\affiliation{School of Astronomy and Space Sciences,
University of Science and Technology of China, Hefei, 230026, China}

\author[0000-0002-2242-1514]{Zhengyan Liu}
\affiliation{Department of Astronomy, University of Science and Technology of China, Hefei, 230026, China; jiazheng@mail.ustc.edu.cn, jnac@ustc.edu.cn}
\affiliation{School of Astronomy and Space Sciences,
University of Science and Technology of China, Hefei, 230026, China}

\author{Junhan Zhao}
\affiliation{Department of Astronomy, University of Science and Technology of China, Hefei, 230026, China; jiazheng@mail.ustc.edu.cn, jnac@ustc.edu.cn}
\affiliation{School of Astronomy and Space Sciences,
University of Science and Technology of China, Hefei, 230026, China}

\author{Ziqing Jia}
\affiliation{Department of Astronomy, University of Science and Technology of China, Hefei, 230026, China; jiazheng@mail.ustc.edu.cn, jnac@ustc.edu.cn}
\affiliation{School of Astronomy and Space Sciences,
University of Science and Technology of China, Hefei, 230026, China}

\author[0000-0003-4700-348X]{Chengyi Wang}
\affiliation{Department of Astronomy, University of Science and Technology of China, Hefei, 230026, China; jiazheng@mail.ustc.edu.cn, jnac@ustc.edu.cn}
\affiliation{School of Astronomy and Space Sciences,
University of Science and Technology of China, Hefei, 230026, China}



\author[0000-0003-4200-4432]{Lulu Fan}
\affiliation{Department of Astronomy, University of Science and Technology of China, Hefei, 230026, China; jiazheng@mail.ustc.edu.cn, jnac@ustc.edu.cn}
\affiliation{School of Astronomy and Space Sciences,
University of Science and Technology of China, Hefei, 230026, China}
\affiliation{Department of Physics and Astronomy, College of Physics, Guizhou University, Guiyang 550025, People's Republic of China}
\affiliation{Institute of Deep Space Sciences, Deep Space Exploration Laboratory, Hefei 230026, China}

\author[0000-0002-7660-2273]{Xu Kong}
\affiliation{Department of Astronomy, University of Science and Technology of China, Hefei, 230026, China; jiazheng@mail.ustc.edu.cn, jnac@ustc.edu.cn}
\affiliation{School of Astronomy and Space Sciences,
University of Science and Technology of China, Hefei, 230026, China}
\affiliation{Institute of Deep Space Sciences, Deep Space Exploration Laboratory, Hefei 230026, China}

\author{Feng Li}
\affiliation{Department of Astronomy, University of Science and Technology of China, Hefei, 230026, China; jiazheng@mail.ustc.edu.cn, jnac@ustc.edu.cn}

\author{Ming Liang}
\affiliation{National Optical Astronomy Observatory (NSF’s National Optical-Infrared Astronomy Research Laboratory) 950 N Cherry Ave. Tucson Arizona 85726, USA}


\author{Jinling Tang}
\affiliation{Institute of Optics and Electronics, Chinese Academy of Sciences, Chengdu 610209, China}

\author{Hairen Wang}
\affiliation{Purple Mountain Observatory, Chinese Academy of Sciences, Nanjing 210023, China}

\author[0000-0003-1617-2002]{Jian Wang}
\affiliation{Department of Astronomy, University of Science and Technology of China, Hefei, 230026, China; jiazheng@mail.ustc.edu.cn, jnac@ustc.edu.cn}
\affiliation{Institute of Deep Space Sciences, Deep Space Exploration Laboratory, Hefei 230026, China}
\affiliation{State Key Laboratory of Particle Detection and Electronics, University of Science and Technology of China, Hefei 230026, China}

\author[0000-0002-1935-8104]{Yongquan Xue}
\affiliation{Department of Astronomy, University of Science and Technology of China, Hefei, 230026, China; jiazheng@mail.ustc.edu.cn, jnac@ustc.edu.cn}
\affiliation{School of Astronomy and Space Sciences,
University of Science and Technology of China, Hefei, 230026, China}

\author{Ji Yang}
\affiliation{Purple Mountain Observatory, Chinese Academy of Sciences, Nanjing 210023, China}

\author[0000-0002-1463-9070]{Hongfei Zhang}
\affiliation{Department of Astronomy, University of Science and Technology of China, Hefei, 230026, China; jiazheng@mail.ustc.edu.cn, jnac@ustc.edu.cn}
\affiliation{State Key Laboratory of Particle Detection and Electronics, University of Science and Technology of China, Hefei 230026, China}

\author[0000-0002-1330-2329]{Wen Zhao}
\affiliation{Department of Astronomy, University of Science and Technology of China, Hefei, 230026, China; jiazheng@mail.ustc.edu.cn, jnac@ustc.edu.cn}
\affiliation{School of Astronomy and Space Sciences,
University of Science and Technology of China, Hefei, 230026, China}


\author[0000-0003-0694-8946]{Qingfeng Zhu}
\affiliation{Department of Astronomy, University of Science and Technology of China, Hefei, 230026, China; jiazheng@mail.ustc.edu.cn, jnac@ustc.edu.cn}
\affiliation{School of Astronomy and Space Sciences,
University of Science and Technology of China, Hefei, 230026, China}


\begin{abstract}
We report the discovery of tidal disruption event (TDE) WFST250820mmsw/AT2025wet by the 2.5-meter Wide Field Survey Telescope (WFST). It exhibits a blue nuclear flare throughout the observed evolution with a $g$-band peak magnitude $\sim22$, which is about 3 magnitudes brighter than its host galaxy. A Keck/LRIS spectrum taken near the optical peak reveals a featureless blue continuum, with no discernible  emission lines. However, its redshift can be accurately determined to be 1.037 by its host galaxy absorption lines. Blackbody fits to the multiband spectral energy distribution (SED) of AT2025wet yield a constant temperature of $\sim19,000$~K and a peak luminosity of $(8.27^{+0.92}_{-0.71})\times10^{44}~\rm erg\,s^{-1}$ while actually the SED likely peaks at a much shorter wavelength than a $19,000$~K blackbody. 
The SED modeling of the host galaxy implies a stellar mass of $\sim10^{11.2}\,M_\odot$ and an estimated central black hole mass of $\sim10^8\,M_\odot$, with no evidence of significant active galactic nucleus activity prior to the flare. All of these observations are well consistent with a featureless TDE scenario, making it the highest-redshift non-jetted TDE known to date. TDEs at such high redshift provide us a unique opportunity to explore the intrinsic SEDs of TDEs, particularly to test whether they peak in the extreme-UV regime, thereby addressing the missing energy puzzle and the origin of optical emission in TDEs. Ongoing surveys represented by WFST and the Legacy Survey of Space and Time (LSST) are expected to discover an increasing number of TDEs at higher redshifts, which will extend our census of SMBHs across redshift space and help unravel the mysteries of optical TDEs through direct probes of their UV emission.

\end{abstract}


\section{Introduction}

Tidal disruption events (TDEs) occur when a star is torn apart by the tidal gravity of a supermassive black hole (SMBH). These events produce luminous multi-wavelength flares that serve as rare probes of accretion physics and the demographics of quiescent SMBHs~\citep{Rees1988}. The majority of TDEs are discovered in the optical band since 2010s~\citep{Gezari2021} thanks to a variety of wide-field time-domain surveys such as Pan-STARRS~\citep{Gezari2012}, PTF~\citep{Arcavi2014}, ASAS-SN~\citep{Holoien2016} and ZTF~\citep{vV2019}. Particularly, ZTF has substantially increased the TDE discovery rate, opening up the era of large-sample studies~\citep{Velzen2021, Hammerstein2023} and enabling detailed demographic study of local SMBHs~\citep{Yao2023}. Despite significant progress in TDE searches, the current sample of optically detected TDEs remains limited. It is largely confined to low redshifts (e.g., $z \lesssim 0.5$ for the ZTF sample,~\citealt{Lin2022b,Yao2023}) and to the bright end of the luminosity function, with very few events fainter than $L_{g} \lesssim 10^{42.5}$~\lum~\citep{Zhu2023}. Detecting fainter and more distant TDEs is therefore essential for conducting a comprehensive census of dormant SMBHs in the universe.

Moreover, TDEs at higher redshifts are valuable for exploring the intrinsic optical-ultraviolet (OUV) spectral energy distributions (SEDs) of TDEs, which is essential for testing different models that explain the optical emission of TDEs. 
 Currently, the SEDs of individual optical TDEs are empirically described by a blackbody model, exhibiting approximately constant temperatures during their temporal evolution. However, the characteristic blackbody temperatures vary across different TDEs, typically spanning $(1-4)\times10^4$~K~\citep{Velzen2021,Hammerstein2023}. Two main models are under discussion: one suggests that UV/optical emission is a result of the reprocessing of X-ray photons in an extended envelope/outflow~\citep{Loeb1997,Strubbe2009,Metzger2016, Roth2016,Dai2018}, while the other proposes that the emissions are caused by shocks generated by the stream-stream collision of the stellar debris~\citep{Piran2015,JiangYF2016}. The reprocessing model predicts that the SED peaks at extreme-UV~\citep{Dai2018} yet is invisible due to the observing window gap and produce the so-called missing energy puzzle~\citep{Lu2018}. High-redshift TDEs offer a unique opportunity since their rest-frame far-UV emission is shifted into observable wavelengths due to cosmological redshift. This provides a direct and efficient method to measure the full SED shape and thereby test the models of TDE optical emission. 

Deeper observations are an eternal pursuit in astronomical surveys and are especially important for capturing rare transients like TDEs by probing larger cosmic volumes. The Wide Field Survey Telescope (WFST; \citealt{WFST}) is optimized for such discoveries, featuring a 6.5\,deg$^2$ field of view, deep limiting magnitudes of $r\sim23$ in single exposures, and high-cadence multi-band imaging over a wide area. These capabilities enable the detection of distant transients that were previously inaccessible to shallower surveys~\citep{Lin2022a,Hu2022}. Here, we report the discovery of AT~2025wet (WFST250820mmsw), currently the highest-redshift optical TDE candidate known ($z=1.037$). AT~2025wet was located in the field of the Deep High-cadence ugr-band Survey project (“DHugr”; J. Jiang et al. 2026, in preparation). The DHugr is a time-domain program designed to exploit WFST’s superior \textit{u}-band imaging capability. It will monitor $\rm \sim2\times360\,deg^{2}$ around the celestial equator each year, covering the Spring and Autumn fields with six months of observations per field. The survey will obtain daily \textit{u}-band photometry, together with at least one additional band, over consecutive $\pm7\,$days intervals in each lunar cycle, while continuing $gr$-band monitoring on the remaining nights. This strategy enables the early detection and characterization of energetic transient phenomena (e.g., \citealt{Wu2026,Zhao2026,Zheng2026}). With an observed peak magnitude of $r\sim22$, corresponding to an absolute magnitude $\rm M_r\sim-22.2$, AT~2025wet is among the most luminous TDEs detected in the optical but the apparently faintest, demonstrating the ability of WFST to probe the TDE population at cosmic noon. 

We describe our discovery, follow-up observations, and data reduction in Section~2, followed by the analysis of its properties and identification as a robust TDE candidate in Section~3. Finally, we briefly discuss our results and conclusions in Sections~4 and 5. For this work, we adopt the cosmological parameters $H_0=70\,\mathrm{km\,s^{-1}\,Mpc^{-1}}$, $\Omega_{\rm M}=0.3$, and $\Omega_{\Lambda}=0.7$.

\begin{figure*}
\begin{minipage}{1\textwidth}
\centering{\includegraphics[angle=0,width=1\textwidth]{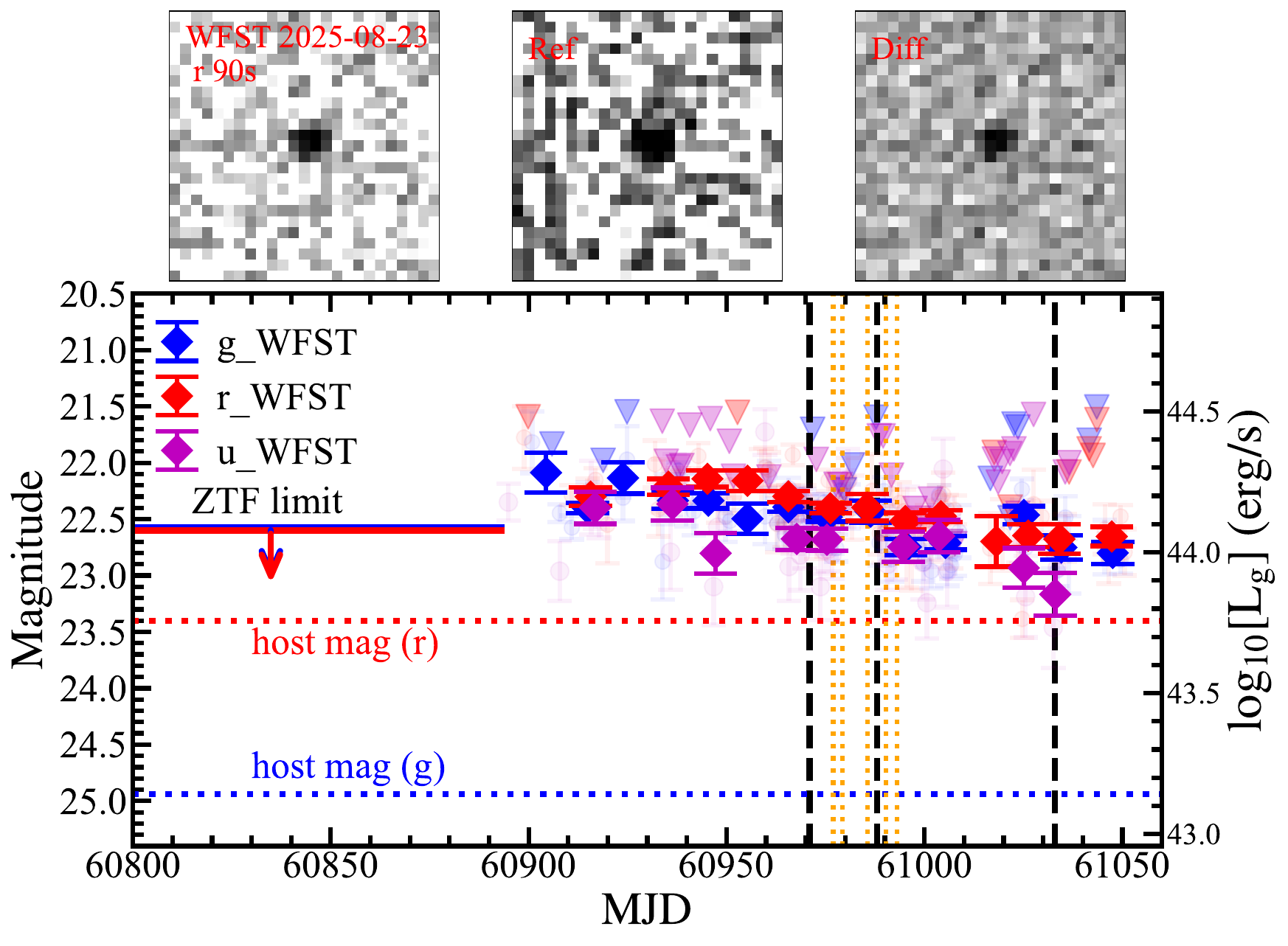}}
\end{minipage}
\caption{Top left: WFST r-band image which was taken on 2025-08-23 UT. Top middle: WFST r-band stack reference image of the host galaxy. Top right: Difference image of AT~2025wet.
Bottom: The multiwavelength light curves of AT~2025wet showing a slow decline with blue color in the optical bands ($\rm \sim100\,days$). The 10-day binned measurements are shown as diamond symbols, overlaid on the individual measurements shown as semi-transparent circles. The black vertical line marks the date of our spectra observation and the triangles represent the upper limits. The orange vertical dotted line marks the date of our Swift observation. The six-month stacked ZTF detection limits are indicated by horizontal lines, while the host-galaxy magnitudes are shown with dashed lines. The data used to create this figure are made available digitally.  \label{LCs}}
\end{figure*}

\section{Observations and Data}

\subsection{Discovery of WFST250820mmsw/AT~2025wet}

This event was initially detected as WFST250820mmsw by WFST on 2025 August 20 and was later reported in the Transient Name Server \citep{25wet_TNS} as AT~2025wet\footnote{\url{https://www.wis-tns.org/object/2025wet}} (hereafter AT~2025wet). The coordinates are $\rm \alpha= 02h15m34.910s$ and $\delta = +01^{\circ}53^\prime30.14^{\prime\prime}$ (J2000) reported by WFST. 
Although this source was only 22.55 mag at discovery, it shows blue color ($g-r=-0.2$) and a slow decline for nearly a month (see Figure~\ref{LCs}). We found the host galaxy of AT~2025wet was detected on \textit{g, r} and \textit{z} band in Legacy survey~\citep{DESI} using the image from The Dark Energy Camera (DECam) and was also detected in Hyper Suprime-Cam Subaru Strategic Program (HSC-SSP, \citet{HSC}). The host galaxy of AT~2025wet is extremely faint in the blue bands, with measured magnitudes of $g = 24.94 \pm 0.08$, $r = 23.38 \pm 0.02$, $i = 21.98 \pm 0.01$, and $z = 21.22 \pm 0.02$, 
resulting in a very red optical color.  Moreover, the Dark Energy Spectroscopic Instrument (DESI) Data Release 1 provides a spectroscopic redshift of $z=1.038$ for the host galaxy, though the spectrum is of low quality with a very low Signal-to-Noise Ratio (SNR).\footnote{\url{https://www.legacysurvey.org/viewer/desi-spectrum/dr1/targetid39627833613489116}} 

AT~2025wet lies at an offset of 
$\Delta \mathrm{RA} = (0.06 \pm 0.09)^{\prime\prime}$ and 
$\Delta \mathrm{Dec} = (0.01 \pm 0.07)^{\prime\prime}$ 
from the host-galaxy centroid reported in the Legacy Survey catalog. 
This corresponds to a 3$\sigma$ upper limit projected physical offset of 
$r < 2.25\,\mathrm{kpc}$ at the redshift of the host.
The host galaxy has a half-light radius of $R_e = 0.245^{\prime\prime}$ in the Legacy Survey DR10 catalog, placing AT~2025wet at a projected offset of $0.24\,R_e$ from the galaxy center, with a $3\sigma$ upper limit of $<1.4\,R_e$. We note, however, that the host is intrinsically compact, and therefore the offset constraint alone does not provide a very strong constraint on the nuclear nature of the transient.

An alternative possibility is that the transient is unrelated to the $z=1.037$ galaxy and instead originates from a foreground or background source projected in close angular alignment. However, the transient position is consistent with the galaxy nucleus within 0.06\,arcsec. Following the standard probability-of-coincidence formalism~\citep{Bloom2002},
\begin{equation}
p_{\rm chance}=1-\exp\left[-\pi r^2 \Sigma(<m)\right],
\end{equation}
where $r$ is the angular offset and $\Sigma(<m)$ is the cumulative sky surface density of galaxies brighter than the host magnitude. Adopting the galaxy number counts from \citet{Hogg1997}, we estimate a chance-coincidence probability of only $p_{\rm chance}\sim5\times10^{-5}$ for our case. Such a low probability indicates that the observed association is unlikely to arise from a random superposition.

Given these considerations, AT~2025wet is likely a powerful high-redshift nuclear blue transient, exhibiting more than a two-magnitude brightening in the $g$ band. The host galaxy spectrum and colors show no indications of active galactic nucleus (AGN) activity. Therefore, despite its faintness, the source clearly merits follow-up observations to better investigate its nature.

\begin{figure*}
\centering
\begin{minipage}{1\textwidth}
\centering{\includegraphics[angle=0,width=1\textwidth]{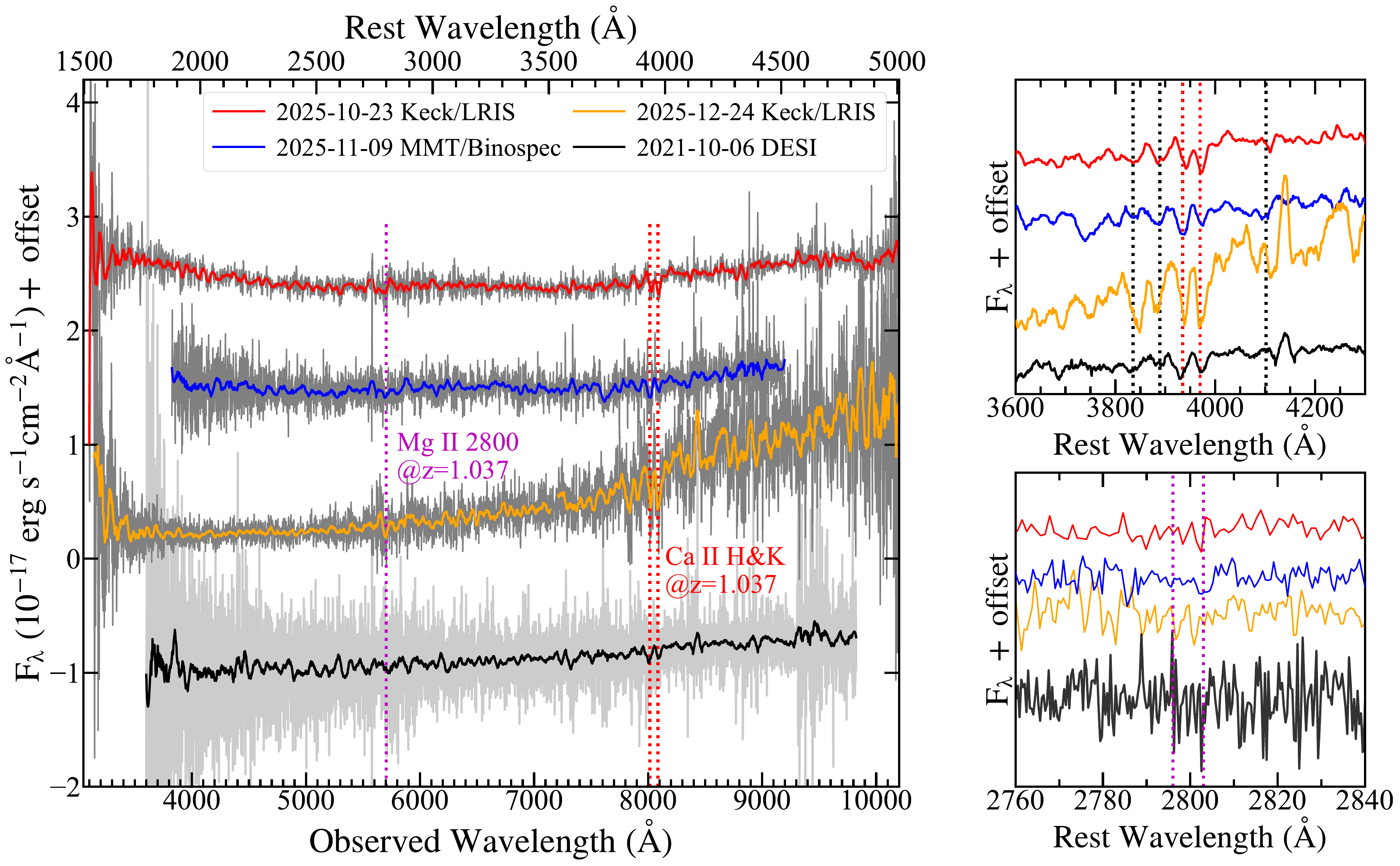}}
\end{minipage}
\caption{Left panel: Optical spectra of AT~2025wet. Each spectrum is labeled by the instrument employed and the date of observation. Smoothed spectra obtained with a Savitzky-Golay filter are shown, with the original, unsmoothed data plotted in gray behind. Ca II $\rm H\&K$ and Mg II $\lambda$2800 absorption lines corresponding to the host-galaxy redshift of $z=1.037$ are marked in the figure. The data used to create this figure are made available digitally.
Right panels: Zoom-in views of the spectral regions around the Ca~II H\&K (red) and Balmer (black) lines in the top panel and the Mg~II $\lambda2796$ \& $\lambda2803$ line (magenta) in the bottom panel. To better display the narrow Mg II lines, we show the unsmoothed spectra in the bottom panel.\label{spec_opt}}
\end{figure*}

\begin{figure*}
\centering
\begin{minipage}{0.95\textwidth}
\centering{\includegraphics[angle=0,width=1\textwidth]{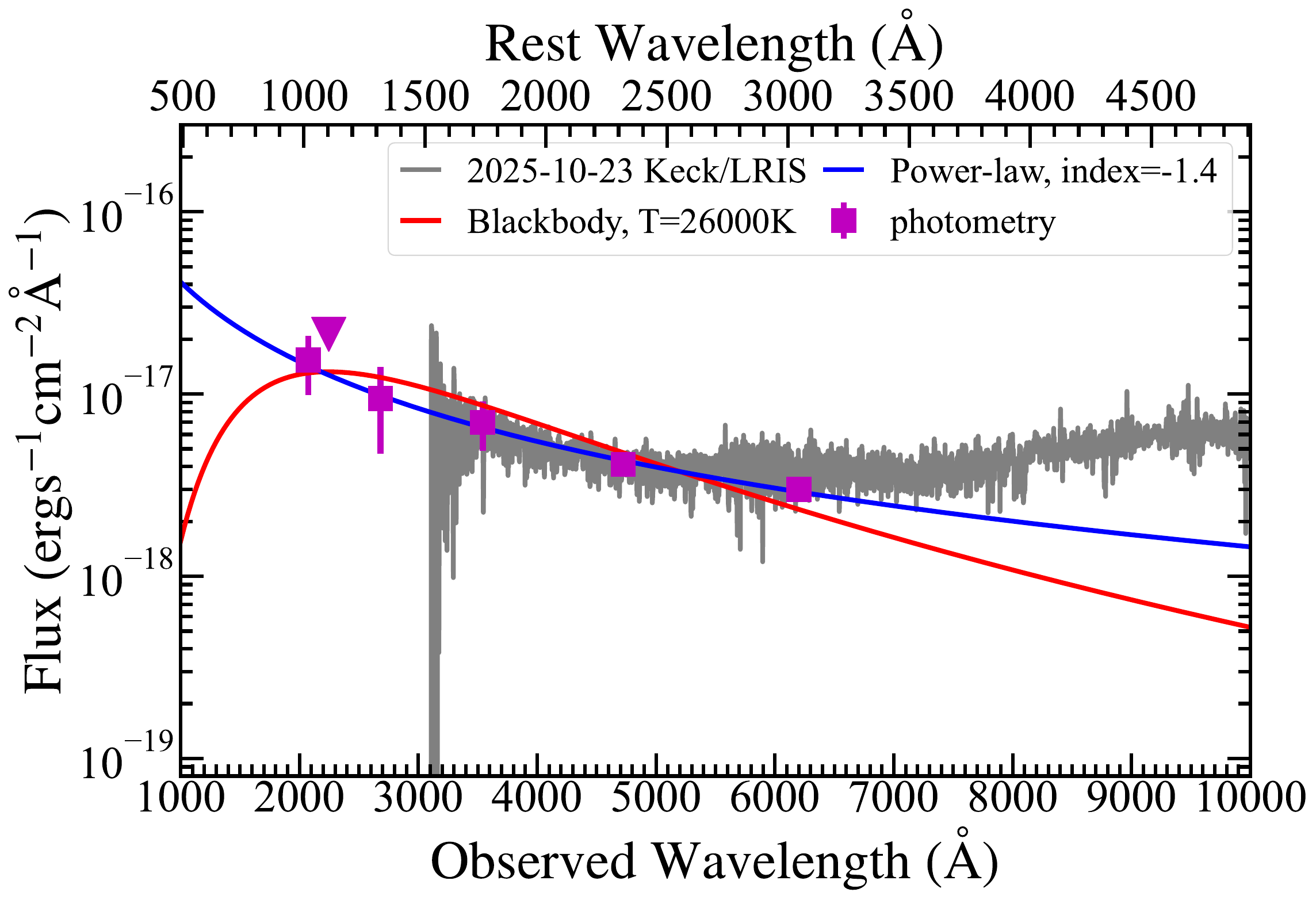}}
\end{minipage}
\caption{Quasi-simultaneous SED fitting of AT~2025wet obtained between 2025 October 23 and November 3. The SED is fitted separately with a power-law model (blue line) and a blackbody model (red line). Our high SNR Keck/LRIS spectrum is overplotted below for comparison.
\label{spec_fit}}
\end{figure*}

\subsection{WFST Data}

AT~2025wet lay within the WFST DHugr footprint, and the imaging data were collected through the survey’s routine observing schedule. The reference image was produced by co-adding individual WFST images to improve depth and SNR. And the photometric measurements were generated automatically by the WFST data reduction pipeline \citep{WFST_pipe}. We utilized a database service, originally developed for variable star studies in WFST observations~\citep{Linjie2025}, which also enables search and filtering of transient sources. Due to the high frequency of observations in the DHugr field, observations are carried out regardless of weather conditions. Therefore, we only selected the data with good conditions for analysis (upper limit $\rm \ge21.5\,mag$), as AT~2025wet is extremely faint.
In Figure~\ref{LCs}, we show the light curve of AT~2025wet binned in 10-day intervals (diamonds), which provides a clearer view of the overall evolution. Individual measurements are overplotted as semi-transparent circles to indicate the full sampling and scatter of the data.

\subsection{Optical Spectroscopic Observation}

We have obtained three optical spectra of AT~2025wet. Two spectra were taken with the Low-Resolution Imaging Spectrometer (LRIS; \citealt{oke95}) on the Keck 10-meter telescope on 2025 October 23 and 2025 December 24. They were reduced using standard tasks following the procedures of \citet{silverman2012}. The total exposure time was 1800 seconds and 3000 seconds respectively, using a 1\arcsec slit and a D560 dichroic to split the light simultaneously into the blue and red arms. The 400/3400 and 600/4000 blue-arm grisms were used in October and December, respectively. Along with the 400/8500 red-arm grating centered at 7865~\AA\, this resulted in a resolving power of R$\sim$1000 and a wavelength coverage of 3100\,\AA\ to 10300\,\AA.
Another spectrum was obtained using the BINOSPEC spectrograph~\citep{Fabricant2019} mounted on the 6.5m Multiple Mirror Telescope (MMT) on 2025 November 9 with exposures of 6000\,s, in which a 270 ($R\sim1400$) grating at a central wavelength of 6500~\AA\ and a 1\arcsec\ long slit were used for the observation. The data was reduced using the standard Binospec IDL pipeline by the Smithsonian Astrophysical Observatory (SAO) staff. All spectra are shown in Figure~\ref{spec_opt}, along with the low quality archival spectrum from DESI.

\subsection{Swift/UVOT photometry}

UV images were obtained with the {\it Neil Gehrels Swift Observatory} (hereafter {\it Swift}) with the Ultra-Violet Optical Telescope (UVOT) from 2025 October 28 to November 14. The {\it Swift} photometry (PI: Jiang) was measured with the UVOTSOURCE task in the {\tt Heasoft} package \citep{heasoft} with the source and background regions defined by circles with radii of $5^{\prime\prime}$ and $30^{\prime\prime}$, respectively. Considering the SED fitting results of the host galaxy in Section~\ref{SUBSECT:HostGal}, the ultraviolet contribution from the host galaxy is completely insignificant in {\it Swift} observation. Thus we try to co-add all the observations to detect the ultraviolet component of AT~2025wet. We detected a signal of $23.07 \pm 0.30\,$mag in a total exposure of 6900\,s in UVW2 band (mean MJD is 60985.0), as well as a marginal $\sim$2$\sigma$ signal of $23.10 \pm 0.47$ mag based on UVW1 image in a total exposure of 5300\,s (mean MJD is 60987.2).

The photometry was calibrated to the AB magnitude system \citep{Gunn1983} and corrected the Galaxy extinction, adopting the revised zero points and sensitivities from \citet{Breeveld2011}.

\subsection{Swift/XRT photometry}
The source was observed with the X-Ray Telescope (XRT) onboard the Neil Gehrels \emph{Swift} Observatory. We processed the data using the HEASoft tasks XRTPIPELINE and XRTPRODUCTS to produce light curves and spectra. Source counts were extracted from a circular aperture of radius $47.1^{\prime\prime}$ centered on the target, while the background was taken from a nearby source-free circular region of radius $150^{\prime\prime}$. No significant X-ray emission was detected in either individual observations or in the stacked images. For non-detections, we computed 3$\sigma$ upper limits on the X-ray flux using the WebPIMMS tool\footnote{https://heasarc.gsfc.nasa.gov/cgi-bin/Tools/w3pimms/w3pimms.pl}, assuming a power-law spectrum with photon index $\Gamma=1.75$ \citep{Ricci2017}. Analysis of the stacked images yields an unabsorbed $3\sigma$ upper limit on the 
X-ray flux of $3.02\times 10^{-14}~\mathrm{erg\,s^{-1}\,cm^{-2}}$, which 
translates to a luminosity of $1.5\times10^{44}~\mathrm{erg\,s^{-1}}$.

\subsection{Einstein Probe Observation}
We also triggered X-ray follow-up observation using Follow-up X-ray Telescope (FXT) onboard the Einstein Probe (EP, \citealt{Yaun2025EP}) on 2025 November 07 UT, with a total exposure time of 2210\,s. We processed the data using the Follow-up X-ray Telescope Data Analysis Software (FXTDAS), employing the FXTCHAIN task to generate light curves and spectra. Source counts were extracted from a circular aperture with a radius of $60^{\prime\prime}$ centered on the target, while the background was taken from an annulus region with inner and outer radii of $100^{\prime\prime}$ and $250^{\prime\prime}$, respectively. Similarly to the XRT observations, no significant X-ray emission was detected. Using the FXT Observation Simulator\footnote{https://epfxt.ihep.ac.cn/simulation}, assuming a power-law spectrum with photon index $\Gamma=1.75$, we obtained an unabsorbed $3\sigma$ upper limit on the X-ray flux of $5.14\times 10^{-14}~\mathrm{erg\,s^{-1}\,cm^{-2}}$, corresponding to a luminosity of $2.6\times10^{44}~\mathrm{erg\,s^{-1}}$.

\subsection{Archival photometry Data}

We also collected host-subtracted light curves of AT~2025wet from public time-domain surveys, the Zwicky Transient Facility (ZTF; \citealt{ZTF}). The ZTF $g-$ and $r-$band light curves were obtained using the ZTF Forced Photometry Service \citep{ZFPS}.  Because single-epoch ZTF images are not sufficiently deep and no useful post-outburst observations are available, we binned all ZTF difference-image measurements from the six months prior to the flare to derive a pre-outburst flux upper limit, following the methodology described in the ZTF forced-photometry guidelines~\citep{ZFPS}.

All light curves, after correction for Galactic extinction, are shown in Figure~\ref{LCs}. We adopted a \citet{Cardelli1989} extinction law with $R_V=3.1$ and a Galactic extinction of $E(B-V)=0.0286\pm0.0004$\,mag (\citealt{Schlafly2011}).

\begin{figure}
\centering
\begin{minipage}{0.5\textwidth}
\centering{\includegraphics[angle=0,width=1\textwidth]{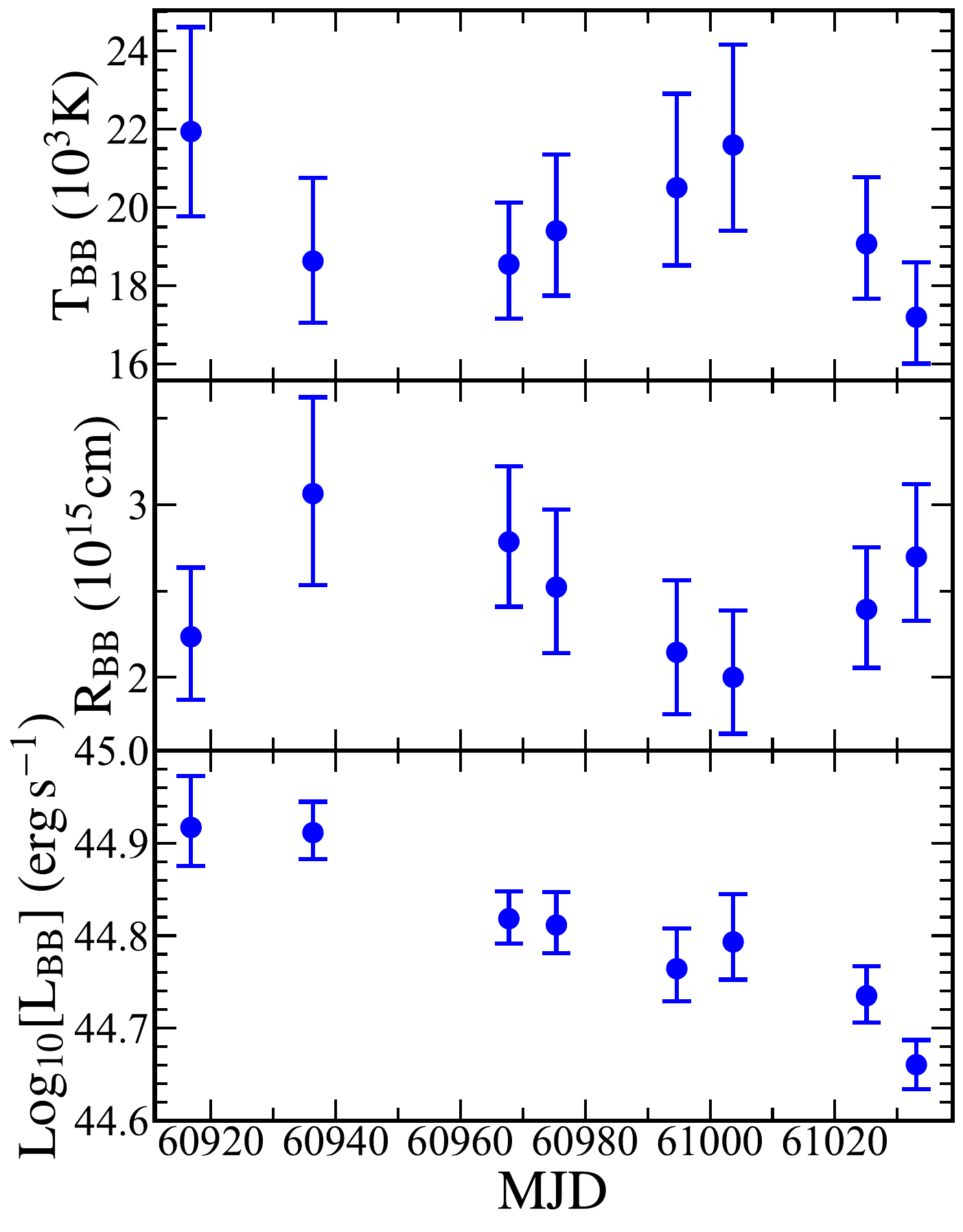}}
\end{minipage}
\caption{The evolution of blackbody temperature, radius and luminosity of AT~2025wet form top to bottom, respectively. The data used to create this figure are made available digitally.
\label{bbfit}}
\end{figure}

\section{Analysis and Results}

\subsection{Spectroscopic Analysis}

We detected the Ca II $\rm H\&K$ and Balmer absorption lines corresponding to the host-galaxy redshift of $z=1.037$ in our spectra and host legacy spectrum. We also examined the expected position of the Mg II $\lambda2796$\,\&\,$\lambda2803$ absorption feature in all epochs (see Figure~\ref{spec_opt}). Although two weak dips is present near the expected wavelength in the 2025-12-24 spectrum, it is offset from the line center. Comparable fluctuations are also seen elsewhere in the spectrum, and no consistent feature is detected in the other epochs. We therefore conclude that the current data do not provide a secure detection of Mg II absorption. In addition, no spectral features indicative of a foreground or higher-redshift background source are detected in any of the available spectra.

The photometric SED of the transient itself of AT~2025wet can be well described by a power-law model, as shown in Figure~\ref{spec_fit}. For comparison, we also show a blackbody fit to this SED with a red solid line, which provides a poorer fit to the data (see Section 3.2 for details of the fitting). Despite the amount of fitted data is limited, the inferred temperature remains high and consistent with the characteristic blackbody properties of featureless TDEs~\citep{Hammerstein2023,Yao2023}, including the recent sources AT2024lhc and AT2024kmq presented by \citet{Yao2026}.
The high-resolution Keck spectrum shows no evidence of AGN-related high-ionization UV lines (such as C IV and C III]) or low-ionization features (including Mg II and \hb ). This absence strongly suggests that the flare is unlikely to be associated with a turn-on AGN. Moreover, given that the Swift UV-band SED follows a natural extension of power-law fitting, the SED of AT2025wet appears to be a largely featureless continuum without prominent metal absorption features. SLSNe-Ic typically exhibit strong UV absorption features in this wavelength range (e.g., \citealt{Quimby2011, Gal-Yam2019}), which are not observed in our spectra. In addition, their host galaxies are predominantly low-mass, star-forming systems, in clear contrast to the host of AT~2025wet, further disfavoring this classification. Thus, we concluded that a featureless TDE at $z=1.037$ provided a plausible explanation.

\subsection{Photometric Analysis}

We model the SED at each epoch by fitting a redshifted blackbody to the multi-band photometry. To construct quasi-simultaneous SEDs without extrapolation, we adopt the binned $u$-band epochs as a common time grid and linearly interpolate the binned $g$- and $r$-band light curves onto these epochs, requiring all interpolated points to lie within the observed ranges and the host-galaxy extinction is assumed to be zero.
We estimate the blackbody parameters using a two-step approach. We first obtain a maximum-likelihood solution via $\chi^2$ minimization with the Nelder--Mead algorithm, and then initialize an MCMC sampler implemented with \texttt{emcee}. We assume uniform priors on physically plausible ranges and adopt a Gaussian likelihood based on the flux residuals.
Parameter uncertainties are derived from the posterior distributions after burn-in, quoted as the median and 16th--84th percentile intervals. The bolometric luminosity is computed for each posterior sample, thereby propagating parameter uncertainties and covariance.
The fit quality is assessed using the minimum and reduced $\chi^2$ values to evaluate the validity of the blackbody model at each epoch.

The blackbody luminosity show a slowly decline while the temperature remain at approximately $\rm19,000\,K$ throughout the observed period. The peak blackbody luminosity is $(8.27^{+0.92}_{-0.71})\times10^{44}~\rm erg\,s^{-1}$. All the results are presented in Figure~\ref{bbfit}. In particular, we construct a quasi-simultaneous SED by combining the {\it Swift}/UVOT ultraviolet and $ugr$ optical photometry obtained near 2025 October 23, and fit both blackbody and power-law model using the MCMC procedure described above. The blackbody fit yields a temperature of $\rm T = 2.63^{+0.14}_{-0.13}\times10^4~{K}$ with a reduced $\chi^2$ of 7.41 for 3 degrees of freedom, indicating a poor representation of the observed SED. In contrast, the power-law model gives a spectral index of $\alpha = -1.40^{+0.02}_{-0.01}$ and a lower reduced $\chi^2$ of 0.63 for 3 degrees of freedom (see Figure~\ref{spec_fit}). Given the limited number of photometric data points, the power-law model appears to provide a more adequate representation of the observed SED of AT~2025wet at this epoch. This also reflect the limitations of a single-temperature blackbody description, which has been discussed in the context of the so-called “missing energy” problem in TDEs (e.g., \citealt{Lu2018}). In this picture, a significant fraction of the expected emission may emerge outside the observed UV/optical bands, and a single-temperature blackbody may therefore not fully capture the total radiative output. Our results are in qualitative agreement with those of \citet{Lin2025}, who found that a power-law model can better describe rest-frame SEDs than a simple blackbody. Considering that the inclusion of the UV data yields a best-fit blackbody temperature of up to $\sim 26,000\,$K, the persistently high temperature observed throughout the lifetime of AT~2025wet remains consistent with those of other featureless TDEs.

\subsection{Host-galaxy Properties}
\label{SUBSECT:HostGal}

\begin{figure}
\centering
\begin{minipage}{0.5\textwidth}
\centering{\includegraphics[angle=0,width=1\textwidth]{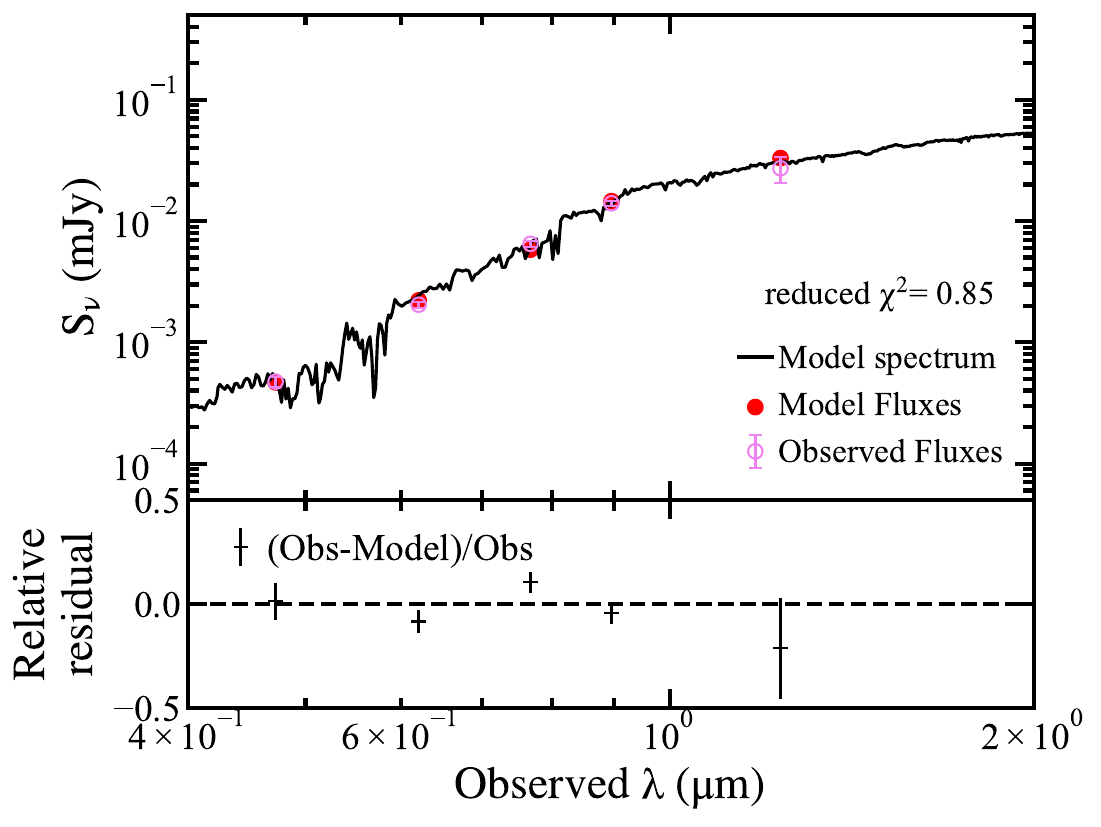}}
\end{minipage}
\caption{The host SED fitting results using the package CIGALE.
    Top panel: The different components considered in the SED fitting. The red points represent the model flux for each band derived from the best-fit SED, and the violet circles indicate the observed data.
    Bottom panel: The residuals between the observed data and model flux.
\label{host_sed}}
\end{figure}

The pre-outburst DESI spectrum of the galaxy nucleus shows no evidence of any AGN component, although the S/N is low.
We collected multi-band photometry of the host galaxy from several archival surveys, including HSC-SSP~\citep{HSC}, SDSS, Legacy, Pan-STARRS and the UKIRT Infrared Deep Sky Survey (UKIDSS, \citet{UKIDSS}). We find no evidence for long-term variability over two decades prior to the outburst, particularly at blue ends, where AT~2025wet is undetected in SDSS and Pan-STARRS, while remaining at $\sim25\,$mag in HSC and DESI. In the $r$ band, all surveys consistently measure a stable magnitude of $\sim23.4\,$mag within the uncertainties.
Thus, we used the Python package Code Investigating GALaxy Emission (CIGALE; \citealt{CIGALE}) to model the SED of the host galaxy, using the most precise optical photometry from HSC and near-infrared measurements from UKIDSS. 
We adopted the same fitting methodology and parameter settings as in \citet{Zhu2025}. 
Our best-fit model, shown in Figure~\ref{host_sed}, yields a reduced $\chi^2$ of 0.85. In our fitting, the stellar mass of the galaxy is $10^{11.25\pm0.18}\,$\msun\ and the star formation rate (SFR) is $\rm log\mathrm{SFR} = -0.98\pm0.65$. Considering the high redshift of AT~2025wet, we use the empirical relation between \mbh\ and the total galaxy stellar mass from \citet{Farrah2025} to estimate a host \mbh\ of $10^{8.11\pm0.58}M_{\odot}$ of AT~2025wet, which finds a bias-corrected relation from massive early-type galaxies (ETGs) hosting AGN at z$\sim$0.8. Additionally, relying on only four photometric measurements is insufficient to robustly assess the presence or absence of AGN component, here we gives an upper limit luminosity of the AGN component of $\rm 1.3\times10^{44}\,erg\,s^{-1}$.

\section{Discussion}

\subsection{Evidence Supporting a High-Redshift TDE for AT~2025wet}

The observed properties of AT~2025wet strongly suggest a TDE origin. Here we summarize the main properties of AT~2025wet as follows.

\begin{enumerate} 

\item[$\bullet$] The Keck/LRIS spectrum taken around the optical peak shows a strong high-temperature continuum without any line features, which is reminiscent of other featureless TDEs. The SED of the transient can be well fitted as a power law component with a index of $-1.4$.

\item[$\bullet$] The peak blackbody luminosity $(8.27^{+0.92}_{-0.71})\times10^{44}~\rm erg\,s^{-1}$ and the temperature remains $\sim$19,000\,K throughout the observed period, which is comparable to featureless TDEs which are most at the high luminosity end of TDEs with high blackbody temperature.

\item[$\bullet$] The stellar mass of the galaxy is $10^{11.18\pm0.12}\,$\msun\ and the estimated host black hole mass of AT~2025wet is $10^{8.03\pm0.55}M_{\odot}$, which is consistent with other optically overluminous TDEs~\citep{Yao2025}.

\item[$\bullet$] No significant X-ray emission was detected in either individual observations or in the stacked images. The 3$\sigma$ upper limits on the X-ray is a luminosity of $1.5\times10^{44}~\mathrm{erg\,s^{-1}}$.

\end{enumerate}

\begin{figure*}[ht]
\centering
\begin{minipage}{1\textwidth}
\centering{\includegraphics[angle=0,width=1\textwidth]{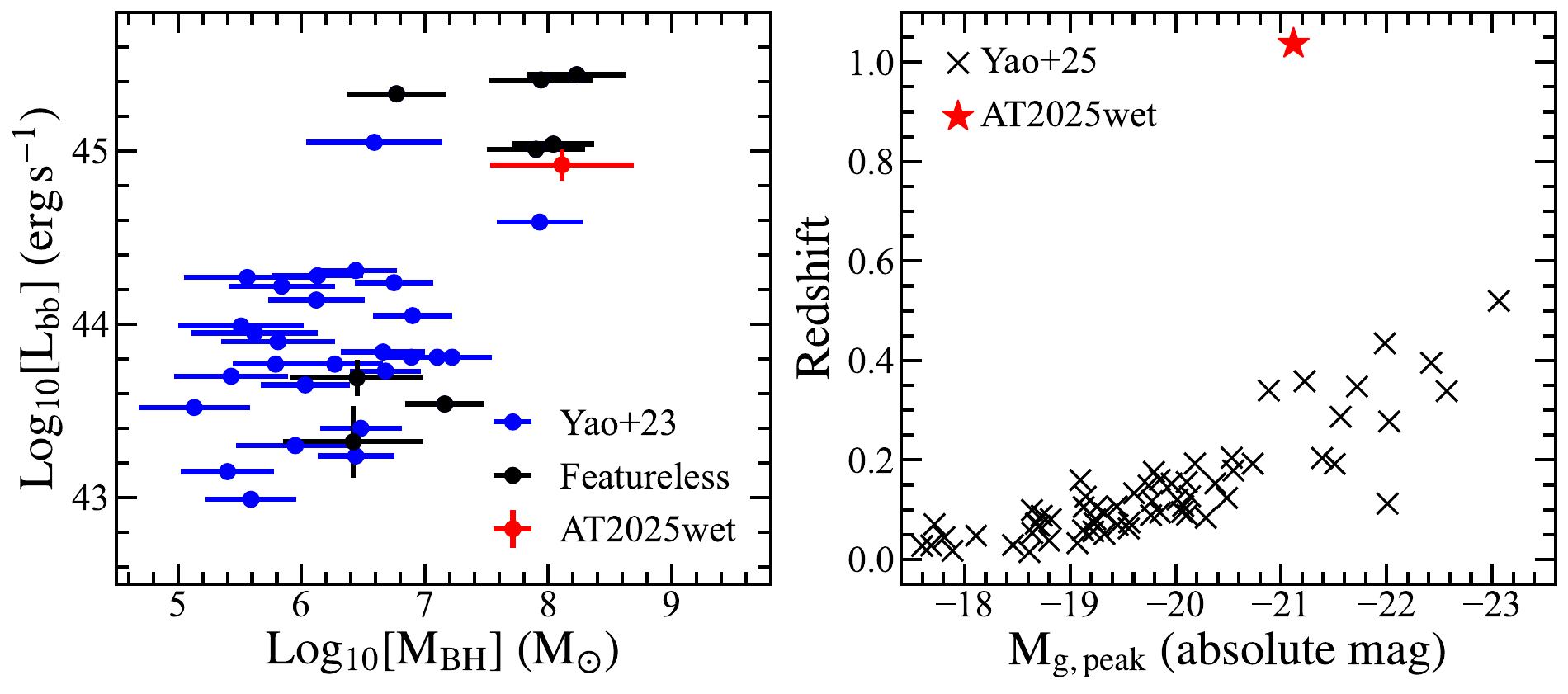}}
\end{minipage}
\caption{Left panel: Peak blackbody luminosity ($L_{\rm BB}$) versus black hole mass (\mbh). We compare AT~2025wet (red dot) with other optical TDEs (blue dots) from \citet{Yao2023}, with the featureless subclass indicated by black dots. Right panel: ZTF-selected TDEs (2019.0–2022.3, \citealt{Yao2025,Yao2026}) on the diagram of redshift versus peak rest-frame g-band absolute magnitude. AT~2025wet are marked by a red star.
\label{compared}}
\end{figure*}

These multi-wavelength properties are difficult to reconcile with expectations for a turn-on AGN. A key discrepancy lies in the persistent absence of broad emission lines, a defining hallmark of turn-on episodes \citep{Gezari2017,Yan2019}. However, there exists a rare subclass of AGNs generally found at high redshifts ($z\gtrsim1.5$), known as weak-emission-line quasars (WLQs; \citealt{McDowell1995,Fan1999}), which exhibit notably weak or absent broad emission lines in their optical (rest-frame UV) spectra. Even so, WLQs generally retain a detectable \mgii$\lambda2799$ emission line \citep{Paul2022}, and near-infrared spectroscopy shows that their \hb\ emission is comparable to that of normal quasars \citep{chen2024}. This behavior is inconsistent with the completely featureless spectra of AT~2025wet.

Although no radio observations are available for AT~2025wet, jetted TDEs are typically characterized by strong X-ray emission~\citep{Levan2011,AT2022cmc}. As an example with well-sampled optical and X-ray observations, AT~2022cmc ($z=1.19325$) had a peak isotropic X–ray luminosity of $\rm L_X\sim3\times10^{47}~erg\,s^{-1}$ and exhibited a $\rm L_{\rm X}/L_{\rm opt,bb} > 30$, which remained in a persistently luminous X-ray state over the optical emission period~\citep{AT2022cmc}. In contrast, AT~2025wet showed a much lower ratio of $\rm L_{\rm X}/L_{\rm bb} < 0.19$, suggesting that it was unlikely to be a jetted TDE.

Time-domain surveys in recent years have also uncovered a wide variety of nuclear transients whose physical origins can be extremely challenging to diagnose. As a result, a number of such events have been grouped into the ambiguous nuclear transient (ANT) class (e.g., \citealt{Neustadt2020,Holoien2022}), most of which occur in AGN environments. Some ANTs reach extremely high luminosities and exhibit line properties resembling those of AGNs (e.g., AT2021lwx; \citealt{Subrayan2023,Wiseman2023}). In contrast, the observed characteristics of AT~2025wet that we summarized above align more closely with those of typical optical, featureless TDEs~\citep{Hammerstein2023,Yao2023}.

\subsection{High-Redshift TDEs as Direct Probes of the UV SED for TDEs}

AT~2025wet provides a rare opportunity to directly probe the UV SED of a TDE. As shown in Figure~3, when extending to wavelengths near $\sim$1000\,\AA, the SED of this featureless TDE becomes increasingly consistent with a power-law form rather than a blackbody. 

Recent studies have highlighted that interpreting the optical blackbody luminosity as the bolometric output of TDEs can be highly inaccurate. \citet{guolo2024} pointed out that the inferred bolometric luminosity depends sensitively on the assumed radiative model; by constructing a disk blackbody SED spanning from optical to X-rays, they emphasized the large uncertainties introduced by the lack of observational constraints in the unobservable extreme-UV band. \citet{Lin2025} found that a power-law model can actually fit the rest-frame $\sim2000-7000$\AA\ SEDs more consistently than a simple blackbody, suggesting that the SEDs should peak at shorter wavelengths. Moreover, direct UV spectral observations by \textit{Hubble Space Telescope}~(HST) of several nearby featureless TDEs, probing down to short wavelengths, further support this conclusion~\citep{Zhu2025}. This indicates that a larger fraction of the radiated energy of optical TDEs is released in the extreme-UV (EUV) i.e., between 100 and 1000~\AA. Parallel evidence comes from dust echo studies of some TDEs, where their extraordinary infrared echoes reveal ultrahigh peak bolometric luminosities~\citep{Jiang2025,Wumx2025}.

If the intrinsic SED peak of optical TDEs is ultimately confirmed to lie in the EUV, this would offer a direct resolution to the missing energy puzzle~\citep{Lu2018}. Moreover, it would also provide compelling evidence in support of the reprocessing model, in which various hydrodynamic simulations consistently predict an SED peak in the EUV regardless of whether they are powered by accretion or shocks~\citep{Dai2018,Thomsen2022,Qiao2025,Huang2026}.
However, the predicted EUV photons from TDEs have long been difficult to detect, yet they may become observable for high‑redshift TDEs. Even for TDEs like AT2025wet at a redshift of $z\sim1$, this raises the possibility that future observations with facilities such as HST could directly detect EUV photons near peak, providing critical insight into the physics of TDE emission and reprocessing. Based on UV spectra of quasars (e.g., \citealt{Shull2025}), we note that neutral hydrogen absorption may not pose a serious issue for TDEs at z$<$2, particularly in host galaxies with relatively low neutral hydrogen column densities. Therefore, for TDEs at $1 \lesssim z \lesssim 2$, and in hosts with modest neutral hydrogen columns, UV SED measurements extending down to $\sim 500$~\AA\ may still be achievable, although this will in general depend on the line-of-sight neutral gas content. In this context, ongoing deep optical time‑domain surveys, such as DHugr of WFST and LSST, will play a key role by systematically discovering and characterizing TDEs at cosmological distances, which remain extremely rare and poorly sampled at present (see right panel of Figure~\ref{compared}). The \textit{James Webb Space Telescope} (JWST) can also play a significant role in this regard. A possible high-redshift TDE candidate has recently been reported based on JWST observations, but its nature remains uncertain due to the lack of spectroscopic redshift measurements and well-sampled light-curve information~\citep{Karmen2025}. In this sense, AT2025wet represents a pioneering high‑redshift TDE with detailed spectroscopic characterization, opening the era of systematic discovery of high‑redshift TDEs and direct probes of their EUV emission.

\section{Conclusion}

Although the optically selected TDE sample has grown rapidly in recent years, these predominantly low-redshift events are intrinsically limited in their ability to reveal the EUV SED of the TDE population. Direct constraints on the far-UV emission, however, have become increasingly important for discriminating different models of TDE radiation.

In this work, we report the discovery of a TDE candidate, AT~2025wet, which has the highest redshift ($z=1.037$) among all optically discovered non-jetted TDEs to date, and it is the first non-jetted optical TDE at $z>1$. The discovery of AT~2025wet opens a new observational window into the high-redshift TDE regime, which has so far remained largely unexplored by optical time-domain surveys. At such redshifts, optical observations naturally probe the rest-frame ultraviolet emission, providing rare access to the UV–optical spectral energy distribution of a TDE at cosmological distances. For AT~2025wet, the observed SED shows a featureless continuum extending into the rest-frame UV, broadly consistent with a power-law shape, suggesting that the total radiative output may substantially exceed that inferred from single-blackbody fits alone. By the way, TDEs occurring in this overluminous parameter space are expected to be associated with black holes of masses frequently exceeding $\sim10^8$\,$M_\odot$~\citep{Yao2025,Yao2026}, highlighting the need for a deeper theoretical understanding of TDE physics in the high-redshift domain. For example, general relativistic effects become increasingly important for disruption by highermass black holes for a sun-like star~\citep{G&R2019,Ryu2023} and some theoretical work suggests that the effective Hills mass increases for more massive stellar progenitors~\citep{Huang&Lu2024,Mummery2024}.

Together with the rarity and observational difficulty of identifying such distant events, AT~2025wet demonstrates both the challenges and the scientific potential of studying high-redshift TDEs. This event highlights how deep optical surveys can begin to directly constrain the ultraviolet emission properties of TDEs, offering new opportunities to test models at cosmic noon.

\acknowledgements
We thank the anonymous referee for their very valuable comments, which have improved the manuscript significantly.
This work is supported by the National Key Research and Development Program of China (2023YFA1608100), the Strategic Priority Research Program of the Chinese Academy of Sciences (XDB0550200), the National Natural Science Foundation of China (grants 12522303,12192221,12393811,12393814), the Fundamental Research Funds for Central Universities (WK2030000097) and the China Manned Space Project.  The authors appreciate the support of the Cyrus Chun Ying Tang Foundations. 
The Wide Field Survey Telescope (WFST) is a joint facility of the University of Science and Technology of China, Purple Mountain Observatory.

Some of the data presented herein were obtained at Keck Observatory, which is a private 501(c)3 non-profit organization operated as a scientific partnership among the California Institute of Technology, the University of California, and the National Aeronautics and Space Administration. The Observatory was made possible by the generous financial support of the W. M. Keck Foundation. 
The authors wish to recognize and acknowledge the very significant cultural role and reverence that the summit of Maunakea has always had within the Native Hawaiian community. We are most fortunate to have the opportunity to conduct observations from this mountain.


\bibliography{AT2025wet.bib}
\bibliographystyle{aasjournal}

\end{document}